\documentclass[fleqn,usenatbib]{mnras}

\usepackage{newtxtext,newtxmath}
\usepackage{threeparttable}
\usepackage{booktabs, caption, makecell}
\usepackage{gensymb}
\usepackage{fix-cm}

\usepackage[T1]{fontenc}
\DeclareRobustCommand{\VAN}[3]{#2}
\let\VANthebibliography\thebibliography
\def\thebibliography{\DeclareRobustCommand{\VAN}[3]{##3}\VANthebibliography}

\usepackage{graphicx}	
\usepackage{amsmath}	
\usepackage{hhline}
\usepackage{multirow}
\usepackage{subcaption}
\usepackage{tabularx}

\title[Radio Emission from a Nearby M dwarf Binary]{Radio Emission from a Nearby M dwarf Binary }

\author[Kelvin Wandia et al.]{Kelvin Wandia,$^{1}$\thanks{E-mail: kelvin.wandia@manchester.ac.uk}
Michael A. Garrett,$^{1,2}$
Robert J. Beswick,$^{1}$
Jack F. Radcliffe,$^{1,3}$
Vishal Gajjar,$^{4,5}$
\newauthor
David Williams-Baldwin,$^{1}$
Chenoa Tremblay,$^{4,10}$
Iain McDonald, $^{1}$
Alex Andersson, $^{6}$
\newauthor
Andrew Siemion,$^{1,4,5,6,7,8,9}$ 
\\
$^{1}$Jodrell Bank Centre for Astrophysics (JBCA), Department of Physics \& Astronomy, Alan Turing Building, The University of Manchester, M13 9PL, UK\\
$^{2}$Leiden Observatory, Leiden University, PO Box 9513, 2300 RA Leiden, The Netherlands\\
$^{3}$Department of Physics, University of Pretoria, Lynnwood Road, Hatfield, Pretoria, 0083, South Africa \\
$^{4}$SETI Institute, 339 Bernardo Ave, Suite 200, Mountain View, CA 94043, USA\\
$^{5}$Berkeley SETI Research Center, University of California, Berkeley, CA 94720, USA \\
$^{6}$Breakthrough Listen, Astrophysics, Department of Physics, The University of Oxford, Keble Road, Oxford, OX1 3RH, UK \\
$^{7}$Astrophysics, Department of Physics, University of Oxford, Keble Road, Oxford, OX1 3RH, UK \\
$^{8}$Berkeley SETI Research Center, University of California, Berkeley, CA 94720, USA \\
$^{9}$University of Malta, Institute of Space Sciences and Astronomy, Msida, MSD2080, Malta \\
$^{10}$Department of Physics and Astronomy, University of New Mexico, Albuquerque, NM 87131, USA\\
\\
}
\date{Accepted XXX. Received YYY; in original form ZZZ}

\pubyear{\the\year{}}

\begin{document}
\label{firstpage}
\pagerange{\pageref{firstpage}--\pageref{lastpage}}
\maketitle

\begin{abstract}
We present the detection of the binary system 2MASS J02132062+3648506 AB using the Karl G. Jansky Very Large Array (VLA) archive data observed at 4--8 GHz. The system is a triple consisting of a tight binary ($\sim0.2\arcsec$) of two M dwarfs of spectral types M4.5 and M6.5 and a wide T3 brown dwarf companion ($\sim$16.4\arcsec). The binary displays coronal and chromospheric activity as traced by previously measured X-ray flux and H$\alpha$ emission. We detect the unresolved binary at a peak flux density of $\sim356\ \mu \mathrm{Jybeam}^{-1}$ at a signal-to-noise ratio (SNR) of $\sim36$ and determine a radio luminosity of $\mathrm{log}L_R/\mathrm{log}L_\mathrm{bol}\sim-7.76$. The radio emission is quiescent, polarised at a mean circular polarisation fraction $f_\mathrm{c}=45.20 \pm 1.58$\% and exhibits a spectral index $\alpha=-0.44\pm0.07$ .  We probe the binary using the Enhanced Multi-Element Remotely Linked Interferometer Network (e-MERLIN) with an angular resolution of $\sim40$ mas at 5\,GHz and detect a component at a peak flux density of $\sim90\ \mu$Jy\,$\mathrm{beam}^{-1}$ at a SNR $\sim5$ . We propose a gyrosynchrotron origin for the radio emission and estimate a magnetic field strength $B<174.86$\,G, an emitting region of size $L<1.54$ times the radius of the M4.5 primary and a plasma number density $n_\mathrm{e}<2.91\times10^5\ \mathrm{cm}^{-3}$. The brown dwarf companion is not detected. Additionally, we have analysed observations of 2MASS J04183483+213127, a chromospherically active L5 brown dwarf  which is also not detected and can only place $3\sigma$ flux density upper limits at $36.9\ \mu$Jy\,$\mathrm{beam}^{-1}$ and $42.3\ \mu$Jy\,$\mathrm{beam}^{-1}$ for Stokes I and V respectively.

%

\end{abstract}

\section{Introduction}

Low-mass main-sequence stars of spectral class M (dwarfs) dominate the stellar content of the Milky Way and constitute $70\mbox{--}75\%$ of the total stellar population \citep[e.g.][]{Henry2006,Winters2019}. Volume-limited surveys for M dwarfs reveal stellar multiplicity and companion rates of $\sim26.8\%$ and $32.4\%$, respectively \cite[see][]{Winters2019}. As stars transit through the M class, their mass and radius decrease, resulting in physical changes to the stars. Besides the directly observable changes e.g. the luminosity, the stellar interior also undergoes a transformation. At a mass $M > 0.35\,M_\odot$, the interior is Solar-like and consists of the core and a radiative and convective zone separated by a tachocline. For stars of lower mass, the radiative zone disappears, resulting in a fully convective interior \citep[see][]{Baraffe2018}. This decrease in mass results in a decrease in pressure on the core and subsequently a decrease in temperature. At a mass $\sim0.075M_\odot\ (78.5\ M_\text{Jup})$ \citep[][]{Chabrier2023} the core temperature is less than the critical temperature of $T>3\times10^{6}$ K \citep[][]{Nelson1993,Burrows1997} required to start and sustain hydrogen fusion and the hydrogen mass burning limit is reached. We note that metallicity plays an important role in the hydrogen mass burning limit. 

Although the existence of very low-mass stars and brown dwarfs was theorised in the 1960s \citep[see][]{Kumar1962}, they remained elusive to observational campaigns. Until the mid 1990s, the coolest conformed main-sequence stars were of spectral type $M9.5\ V$ leading to the view that the main-sequence ended there. This idea was revised after the discovery of sub-stellar objects by \cite{Kirkpatrick1999} using the 2-Micron All-Sky Survey \citep[2MASS; ][]{Skrutskie2006}. The effective temperatures of these objects did not align with the original stellar classification, and they were assigned the spectral class L and T. Most of the objects in the new classes, along with some late-type M dwarfs,  are brown dwarfs. Objects of spectral types $\geq$~M7~V are commonly referred to as ultra-cool dwarfs \citep[][]{Cushing2006}.. We note that brown dwarfs are physically defined by mass rather than the effective temperatures \citep[][]{Reiners2007}. A common method to assess the substellarity of young brown dwarfs is the so-called Lithium test. This test is based on the premise of efficient mixing in the object's convective layers and the core never reaching the ignition temperature of Lithium at $\sim2.5\times10^6$ K, which is lower than that of hydrogen. Consequently, any detection of the Lithium 1 resonance feature at $6708$\ \r{A} indicates the object is unable to fuse hydrogen \citep[e.g.][]{Rebolo1996,Basri1996} and is potentially substellar. 

Chromospheric and coronal activity is routinely observed in many stars. Common tracers of activity in low-mass stars are the first hydrogen Balmer line, H$\alpha$, emitted at optical wavelengths from chromospheric gas heated to $\sim10^4$ K and X-ray fluxes produced by plasma heated to $\sim10^{6}$ K in the corona. This activity is driven by magnetic fields believed to be generated by powerful dynamos at work in the stars. For Solar-like stars with a radiative core and a convective envelope, an $\alpha\Omega$ dynamo \citep[][]{Parker1955} is at work whilst for fully convective stars, an $\alpha^2$ dynamo has been proposed. Although the exact dynamo mechanism is still under investigation, \cite{Chabrier2006,Dobler2006,Browning2008} have shown that the $\alpha^2$ dynamo is able to generate and sustain large-scale magnetic fields.

A critical component of dynamo theory is rotation, which presents an interesting interplay with activity. Activity is observed to increase with rotation for stars ranging from mid-F spectral types to M8 \citep[see][]{Mohanty2002} and peaks in L0 dwarfs with $\sim90\%$ of objects in this spectral type showing activity, then decreases to $50\%$ for L5 dwarfs and becomes increasingly rare in later type objects \citep[][]{Schmidt2015}. The relationship between the rotation and activity has led to the formulation of the rotation-activity paradigm \citep[see][]{Reiners2012}. Although activity increases with rotation it is ultimately constrained by the saturation limit. The saturation limit is the ratio of the corresponding X-ray luminosity $L_\text{X}$ or the $\text{H}\alpha$  luminosity $L_{\text{H}\alpha}$ to the bolometric luminosity and is approximately $\text{log}L_\text{X} / \text{log}L_\text{bol}\approx-3$ \citep[see][]{Vilhu1984} and  $\text{log} \ L_{\text{H}\alpha}/\text{log}\ L_{\text{bol}}\approx-3.8$ for most M dwarfs. Indeed, \cite{Newton2017} have demonstrated that rapid rotators have $L_{\text{H}\alpha}/L_{\text{bol}}$ close to the saturation limit. In regards to binaries, tidal interactions and the exchange of angular momentum increases the rotation in close binaries (<7\arcsec) and, in turn, the chromospheric emission \textcolor{black}{\citep{Hawley1996,Morgan2012}.}

Particle acceleration mechanisms mediated by magnetic fields lead to coherent and incoherent emissions at radio frequencies. Incoherent emission is produced through several radiation mechanisms: thermal bremsstrahlung generated by free electrons in the heated bulk plasma, gyroemission arising from thermal electrons, and gyrosynchrotron originating from non-thermal electrons accelerated along magnetic field lines to mildly relativistic velocities \citep[see][and references therein]{Nindos2020}. Coherent emission is primarily from plasma emission \citep[see][]{Melrose1980} and the electron cyclotron maser \citep[see][]{Wu1979,Melrose1982,Melrose2017}. Following the rotation-activity relationship, a radio activity-rotation relationship ensues. Under this paradigm radio emission from stars of spectral type M0--M6  with rotational velocities $>5\ \text{km\,s}^{-1}$  increases with rotation and saturates at \textcolor{black}{a ratio of the corresponding radio luminosity $L_\text{R}$} to the bolometric luminosity at $\text{log} L_\text{R}/\text{log}L_\text{bol}\approx10^{-7.5}$. For later spectral types, $\text{log} L_\text{R}/\text{log}L_\text{bol}$ is independent of rotation \citep[][]{McLean2011}. A partially analogous relationship is the G\"{u}del-Benz relationship between soft X-ray and radio luminosity at 5 GHz given as $\text{log} L_\text{X} \lesssim \text{log} L_\text{R}\ +15.5$ \citep[][]{GudelBenz1993}. This relationship is valid for all active stars but fails to explain the radio emission observed from ultra cool dwarfs \citep[see][]{Berger2002}. This is conceivably a consequence of saturation in X-ray emission and an unimpeded increase in the radio emission for rapidly rotating ultra cool dwarfs. We note in passing that rapid rotation is not the sole condition required for radio emission as demonstrated by the null detection of rapidly rotating low-mass stars by \cite{Antonova2013}. 

Motivated by the need to better understand the nature of the radio emission from ultra-cool dwarfs, we undertook an analysis of archival National Science Foundation Karl G. Jansky Very Large Array (VLA) data sets targeting the then recently discovered UCDs 2MASS J02132062+3648506 C of spectral type T3 and  2MASS J04183483+2131275 of spectral type L5. The T3 UCD is the tertiary component in an orbit around an X-ray active binary M dwarf and the L5 UCD is chromospherically active and a member of the Hyades cluster. Our aim was to characterise the radio properties and assess the implications of non-thermal emission mechanisms from the UCDs and the binary. In the sections that follow, we describe the targets and their relevant background in Section 2. Section 3 outlines the observational setup and data analysis procedures. We present our results and interpretation in Section 4 and conclude with a summary of our findings and their significance in Section 5.

\section{Targets}
\subsection{2MASS J0213+3648 ABC}

The M dwarf binary 2MASS J02132062+3648506 AB (hereinafter 2M0213 AB) was first detected by \cite{Riaz2006} and reported as a single star of spectral type M4.5. The star displayed significant X-ray emission at $\text{log}\left(L_\text{X}/L_\text{bol}\right) = -3.16$ close to the saturation limit of $-3$. Further follow-up by \cite{Janson2012} characterised the star as a tight binary consisting of a primary and a secondary of spectral types M4.5 and M6.5 respectively and separated by $\sim0.217$\arcsec \citep[][]{Janson2014}. The binary displays chromospheric activity as revealed by the wide equivalent widths of the H$\alpha$ emission line at
at $6.6-8.1$ \r{A} \citep[][]{Riaz2006,Deacon2017} and a H$\alpha$ to bolometric luminosity ratio of $\text{log}\left(L_{\text{H}\alpha}/L_\text{bol}\right) = -3.4$ \citep{Deacon2017} which exceeds values typical for M dwarfs at $\text{log}\left(L_{\text{H}\alpha}/L_\text{bol}\right) = -3.8$  \citep[][]{Czesla2008}. \cite{Deacon2017} demonstrated that it is not physically possible for all the  H$\alpha$ emission to be emitted by the M6.5 dwarf, indicating the M4.5 primary is also chromospherically active. We emphasise that similar statements were not made for X-ray fluxes. 2M0213 AB is a rapid rotator, as demonstrated by the large projected rotational velocity $v\sin{i}\sim25.1$ \citep[][]{Bowler2023}. Considering the two components of 2M0213 AB are of spectral types M4.5 and M6.5, they are fully convective \citep{Stassun2011} and are expected to have efficient dynamos capable of generating kilo-gauss magnetic fields \citep[see][]{Reiners2012}. The binary has a wide ($\sim16.4$\arcsec, $\sim234$ AU) brown dwarf companion of stellar type T3. Evolutionary models have placed the mass and effective temperature at $68\pm7\ M_\text{Jup}$ and $1641\pm167$ K respectively \citep[][]{Deacon2017}. The system is at a distance of $14.28$ pc from the Sun \citep[][]{Lindegren2021}. See Table~\ref{tab:physical_parameters} for the properties of the system.

\subsection{2MASS J0418+2131}

2MASS J04183483+2131275 (hereinafter 2M0418) is a brown dwarf of spectral type L5 first reported by \cite{Garrido2017}. The brown dwarf is a member of the Hyades open cluster, the closest cluster to the Sun and has an age of $500\mbox{--}700$ Myr \citep[][]{Garrido2017} and is at a distance of $40.6\pm2.7$ pc. The L5 brown dwarf displays chromospheric activity as revealed by persistent H$\alpha$ emission \citep[][]{Lodieu2018} at $\text{log}_{10}L_{\text{H}\alpha}/\text{log}_{10}L_{\text{bol}}\sim-6.0$ \citep[][]{Garrido2017}. Spectroscopic follow-up by \cite{Lodieu2018} has led to the detection of a Lithium feature at $6708$\,\r{A}, confirming the sub-stellarity of the object and placing mass upper limits at $<60\,M_\text{Jup}$ \citep[][]{Lodieu2018}. The properties of the brown dwarf are listed in Table~\ref{tab:physical_parameters}.

\begin{table*}
  \centering
  \renewcommand{\arraystretch}{1.2}
  \begin{tabular}{ccccc}
    \hline
    \multirow{2}{2cm}{Property} & \multicolumn{3}{c}{2MASS J0213+3648 ABC} & 2MASS J0418+2131\\
    & 2MASS J0213+3648 A & 2MASS J0213+3648 B & 2MASS J0213+3648 C & \\
    \hline
    Spectral Type & M4.5 $^a$ & M6.5 $^a$ & T3 $^f$ & L$5$ $^i$  \\  
      Position & \multicolumn{2}{c} { 02h13m20.6754 +36d48m51.5425 $^b$} &02h13m19.8876 +36d48m38.3116 $^k$ & 04h18m34.9970s +21d31m26.63754s $^l$ \\
    $\mu_\alpha$ (mas/yr & \multicolumn{3}{c}{$31.1\pm3.6\ ^k$  } & $141.2\pm4.3\ ^j$   \\
    $\mu_\delta$ (mas/yr) & \multicolumn{3}{c}{$50.1\pm3.7\ ^k$  } & $-51.8\pm4.0 ^j$ \\
    $\bar{\omega}$ (mas)  &  \multicolumn{3}{c}{$70.02\pm0.20\ ^{b,d}$} &   $25.8\pm2.9\ ^j$ \\
    epoch & \multicolumn{2}{c}{2015.5}  & 2014.12.14$^k$ & 2015.4 $^l$  \\
    Separation & \multicolumn{2}{c}{0.217\arcsec\ $^c$} & 16.4\arcsec & \\
    Mass & $0.26\pm0.06 M_\odot\ ^c$ & $0.09\pm0.03M_\odot\ ^c$ &  $68\pm7\ M_\text{Jup}\ ^f$ & $<60\ M_\text{Jup}\ ^h$ \\
    $\text{log}_{10}\left(L_\text{X}/L_\text{bol}\right)$  &\multicolumn{2}{c}{-3.16$^e$} & \\
    $v\ \sin{i}\ (\text{km s}^{-1})$ & \multicolumn{2}{c}{25.1 $^g$} &  \\
    $T_\text{eff}$ & & & $1641\pm167^{}\ ^f$ & $1581\pm113\ ^h$ \\
    $\text{log}_{10}\ \left(L_{H\alpha}/L_\text{bol}\right)$ & \multicolumn{2}{c}{-3.4$^f$} & & -6.0$\ ^i$\\
    \hline
  \end{tabular}

  \caption{Physical parameters of the 2MASS J0213+3648 ABC system and 2MASS J0418+2131. References: $^a$ \protect\cite{Janson2012}, $^b$\protect\cite{GaiaDR3Astrometry} , $^c$\protect\cite{Janson2014} , $^d$ \protect\cite{CNGS2023}, $^e$ \protect\cite{Riaz2006}, $^f$ \protect\cite{Deacon2017}, $^g$ \protect\cite{Bowler2023}, $^h$ \protect\cite{Lodieu2018},  $^i$\protect\cite{Garrido2017}, $^j$\protect\cite{Lodieu2019}, $^k$\protect\cite{Best2020}, $^l$\protect\cite{Marocco2021}. 
  We highlight the discrepancy between the effective temperature $T_\text{eff}$ and mass of the L5 and T3 object. $T_\text{eff}$ and mass seem to increase as one progresses to a later spectral type. It is essential to recognise that these parameters are derived from different stellar evolutionary models, which are sensitive to initial conditions and have slightly different input physics.
  }
\label{tab:physical_parameters}
\end{table*}

\section{Methods}
\subsection{VLA Observations}

We have used unpublished (to the best of our knowledge) archival VLA observations (PI: Jan Forbrich) conducted on 2017-11-15 in response to the detection of a nearby T3 brown dwarf companion to  2M0213 AB by \cite{Deacon2017}. Observations were also conducted for 2M0418, a chromospherically active L5 brown dwarf detected by \citet[][]{Garrido2017}. The observations utilised the B configuration of the VLA at C band (4-8 GHz). The observing bandwidth was divided into 32 spectral windows (spws).The standard VLA calibrator 3C286 was observed for flux and bandpass calibration and J0204+3649 was observed as an amplitude and phase calibrator. The pointing centre was based on the position of 2M0213 AB at coordinates reported by \citet[][]{Janson2012}. The coordinates are at epoch 2007-06-01. Similarly, the pointing centre for 2M0418 was based on coordinates reported in the original 2MASS catalogue \citep[see][]{Cutri2003}. The proper motion of the two targets was not considered. The full details of the proper motion effects on the astrometry at the VLA epoch of observation are discussed in section~~\ref{section:astrometry}. The targets were observed over twelve scans each $\sim4$ minutes and 36 seconds long, yielding a total on-target time of $\sim55$ minutes. The data were recorded in full polarisation mode.

\subsection{e-MERLIN Observations}

Follow-up observations of the 2M0213 AB were conducted using the seven-element (six telescopes are often used for regular observations) Electronic Multi Element Remotely Linked Interferometer Network \citep[e-MERLIN; ][]{Garrington2004} at C band (4.82--5.33 GHz) using five antennas over a 24-hour period from 2025-02-24 to 2025-02-25. Three calibrators were observed (PI Wandia): a flux calibrator 3C286 (1331+3030), a bandpass calibrator OQ208 (1407+2827), which is also useful for polarisation leakage calibration and the phase calibrator 0213+3652 located $\sim0.11\degree$ from the target. During the observations, one antenna failed, resulting in a five-telescope experiment. The data were also heavily flagged due to radio interference leading to noise levels far beyond the expected limits of a six-antenna interference-free data. Consequently, re-observations using the same observation setup were requested and conducted \textcolor{black}{intermittently} using six telescopes over a 48-hour period from 2025-03-13 to 2025-03-14. \textcolor{black}{The target was observed over 173 scans at a median scan duration of $\sim5$ minutes and 56 seconds, yielding a total on-target time of $\sim17.1$ hours.}
The data were recorded at full polarisation. 
 
\subsection{Data Analysis}
\label{section:data_analysis}
We have processed the VLA data using two different versions of the VLA pipeline, which is based on the Common Astronomy Software Applications \citep[CASA;][]{CASA2022}. We have used \textsc{CASA} 5.1.0-74, the recommended version and the recent version of the pipeline, which is based on CASA 6.6.1-17 \footnote{\url{https://science.nrao.edu/facilities/vla/data-processing/pipeline}}. We found the results comparable and opted to use CASA 6.6.1-17. The pipeline exports the data from a Science Data Model-Binary Data Format [SDM-BDF]) file to a measurement set, performs Hanning smoothing, antenna position corrections, gain and amplitude calibration among other VLA specific calibration procedures. The data are also automatically flagged at various stages and the end products are science ready visibilities stored in a measurement set. A description of the CASA pipeline stages is found in \cite{Kent2020}. We have made \textcolor{black}{deconvolved} images of the calibrated visibilities using the CASA task \texttt{tclean} to produce continuum images spanning the entire bandwidth \textcolor{black}{using} the multi-term multi-frequency \citep[MTMFS; ][]{Rau2011} deconvolver, which is native to \texttt{tclean}, to account for the wide bandwidth. All images are Briggs weighted \citep{Briggs1995} with a robust parameter of 0.5.

To measure the spectral index, we \textcolor{black}{split the bandwidth into eight chunks each 500 MHz in size and image them individually}. We then mask out any emission above \textcolor{black}{$5\sigma_\text{t}$} where $\sigma_\text{t}$ is the thermal noise in the image and constrain the emission about the known position using a rectangular box.  For each of the images, the integrated flux and the error is then extracted by fitting an elliptical Gaussian using the CASA task \textcolor{black}{$\texttt{imfit}$}. Finally, we use bootstrapping to resample the fluxes and their associated errors and fit a line of best fit. The spectral index is then determined from the slope of the fitted line. \textcolor{black}{To prepare the data for variability analysis, we first mask the target position, model the background sources within the primary beam and shift the pointing centre of the background subtracted visibilities to the proper motion corrected positions of the target using the CASA task \texttt{phaseshift}. Using the CASA toolkit \texttt{casatools}\footnote{\url{https://casadocs.readthedocs.io/en/stable/api/casatools.html}} \texttt{table} and \texttt{ms} tools to parse the visibilities, we average all frequency channels, spectral windows, and baselines to obtain a single visibility per integration time for correlations \texttt{RR} and \texttt{LL}. We note that analysis in the visibility domain circumvents challenges associated with synthesis imaging and deconvolution e.g. imaging artifacts. The noise properties are also well characterised in the visibility domain making it easier to identify calibration and systematic errors. As a result, analysis in the visibility domain yields more reliable estimates. We follow a similar approach in processing the visibilities for the observation of 2M0418. }

The e-MERLIN data were processed using the e-MERLIN CASA pipeline \citep[eMCP;][]{Moldon2018} based on CASA version 5.8.0 \footnote{\url{https://github.com/e-merlin/eMERLIN_CASA_pipeline}}. The pipeline first imports the fits files to  a measurement set. The data are then flagged to remove radio frequency interference, flux, bandpass and gain calibration are performed and the data are flagged again. The fully calibrated data are then split into individual fields and imaged using \texttt{tclean}. A similar weighting scheme and robust parameter as used for the VLA observations are applied.

\section{Results and Discussion}

\subsection{Astrometry}
\label{section:astrometry}

2M0213 AB is detected by Gaia and has been catalogued in the second  \citep[Gaia DR2;][]{GaiaDR2_summary} and the third \citep[Gaia DR3;][]{Gaia_summary_2023} data releases (DR) but remains unresolved in the two catalogues. The wide T3 companion is below the detection threshold. To assess the quality of astrometry, we analyse four goodness of fit statistics:
the
\texttt{astrometric\_excess\_noise} ($\epsilon_i$), \texttt{astrometric\allowbreak\_excess\allowbreak\_noise\allowbreak\_sig} ($D$), \texttt{astrometric\_params\_solved} and the \texttt{renormalised unit weight error (RUWE)}. $\epsilon_i$ measures the noise introduced due to the discrepancy between the observed source's astrometry and Gaia's astrometric model used for the fit. $D$ is the significance of $\epsilon_i$ with $D>2$ indicating significance \citep[see][]{Lindegren2012}. We note that in Gaia DR2, $\epsilon_i$ is potentially inflated due to the inclusion of modelling errors. Significant values of $\epsilon_i$ have been proposed as a signature of unresolved binaries due to orbital wobbles of the components \citep[see][]{Ghandi2022}. The \texttt{astrometric\_params\_solved} lists the astrometric parameters that have been solved for. A good astrometric solution typically returns the six main astrometric parameters (position in right ascension $\alpha$ and declination $\delta$, associated proper motions in right ascension $\mu_\alpha$ and declination $\mu_\delta$, the parallax $\bar{\omega}$  and the radial velocity $v_\text{r}$). Finally, the \texttt{RUWE} indicates the deviation of the astrometric fit from the observed data \citep[][]{GaiaDR3Astrometry} and has been identified as a good indicator of stellar multiplicity \citep[e.g.][]{Belokoruv2020,Castro2024}. A guiding heuristic for a reliable measurement is $\texttt{RUWE}< 1.4$. Larger values are broadly regarded as an indication of poor measurement and or binarity/multiplicity. Despite the chances that larger \texttt{RUWE} arises from poor measurements it enhances the understanding of an object, especially when combined with other goodness-of-fit parameters.

2M0213 AB has the following five astrometric parameters solved for and catalogued in Gaia DR2, $\alpha$, $\delta$, $\mu_\alpha$, $\mu_\delta$ and $\bar{\omega}$. Gaia DR3 has only three catalogued parameters $\alpha$, $\delta$ and $v_\text{r}$. As a result, we opt to use the astrometric solutions of the former.
The binary has $D\sim145$, which notably indicates that $\epsilon_i$ is significant. As \texttt{RUWE} values are not estimated in the Gaia DR2 catalogue; we use \texttt{textsc{gaiadr2-ruwe-tools}} \footnote{\url{https://github.com/agabrown/gaiadr2-ruwe-tools/tree/master}} to determine \texttt{RUWE} and obtain a value of $\sim3$. The large \texttt{RUWE} coupled with the significant $\epsilon_i$ is indicative of binarity. The poor astrometry, especially in Gaia DR3, could be interpreted in the context of a slightly resolved binary, leading to poor astrometric fitting by the Gaia pipeline. We however caution that the possibility of the measurement being unreliable cannot be completely dismissed. Further astrometry is obtained from a volume-limited survey of ultra-cool dwarfs within 25 pc using the United Kingdom Infra-Red Telescope / Wide Field Camera (UKIRT/WFCAM) by \citet[][]{Best2020}, which has targeted 2M0213 ABC and, in particular, the low-mass T3 component. The UKIRT/WFCAM parallaxes for 2M0213 AB are in agreement with those from Gaia DR2, while the measured proper motions of 2M0213 C at $\mu_\alpha\sim31.1\pm3.6$ $\text{mas\ yr}^{-1}$ and $\mu_\delta=50.1\pm3.7$ $\text{mas\ yr}^{-1}$ significantly differ from those of 2M0213 AB measured by Gaia at $\mu_\alpha=65.38\pm0.46$ $\text{mas\ yr}^{-1}$  and $\mu_\delta=64.89\pm0.38$ $\text{mas\ yr}^{-1}$ . Since the T3 brown dwarf is comoving with 2M0213 AB, the components in the system share the same proper motion. We note that the proper motions for 2M0213 ABC from \citet{Best2020}, which we adopt for our analysis, are also in agreement with proper motions from \cite{Lepine2011} at $\mu_\alpha=24$ $\text{mas\ yr}^{-1}$  and $\mu_\delta=47$ $\text{mas\ yr}^{-1}$ .

\begin{figure*}
  \centering
    \setlength{\tabcolsep}{-20pt} 
    \renewcommand{\arraystretch}{0} 
  \begin{tabular}{ccc}
      \includegraphics[width=0.8\columnwidth]{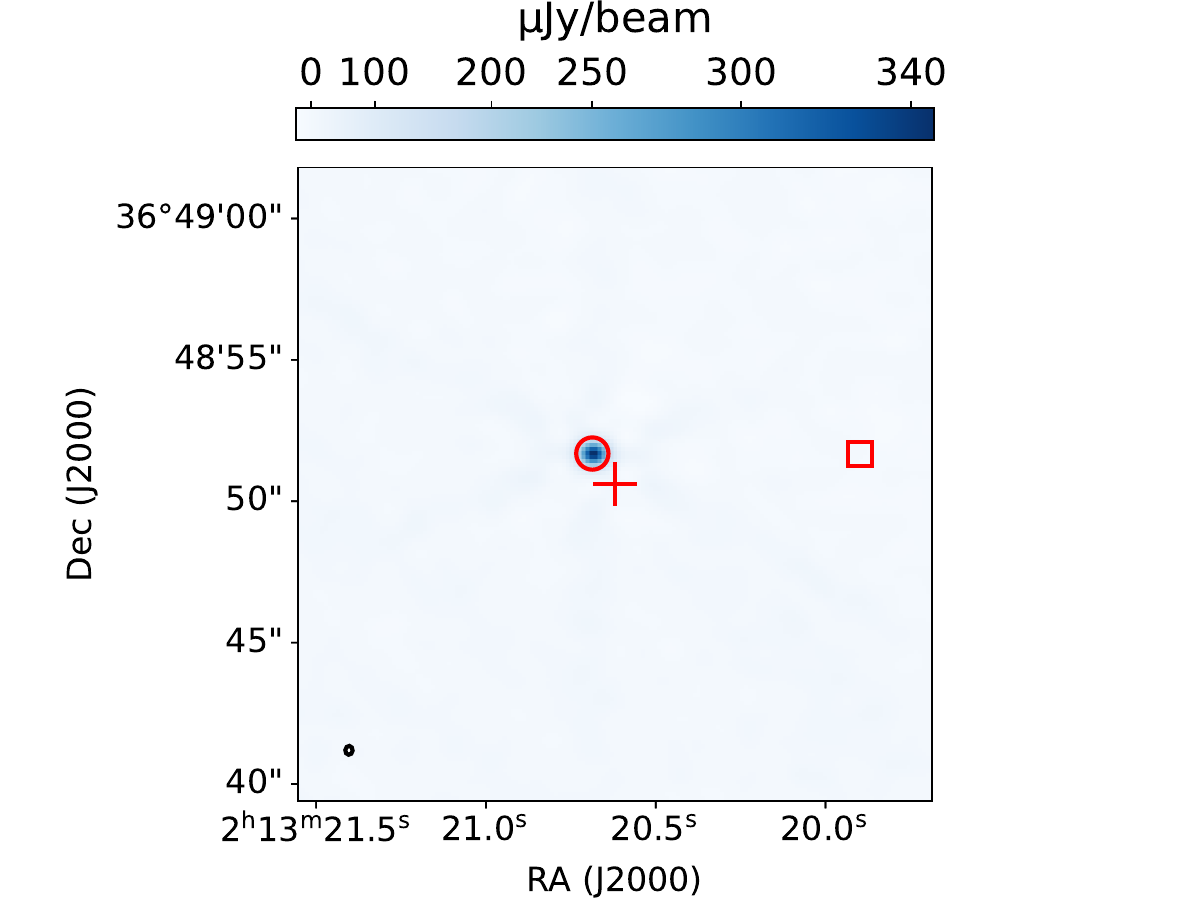} 
      &
      \includegraphics[width=0.8 \columnwidth]{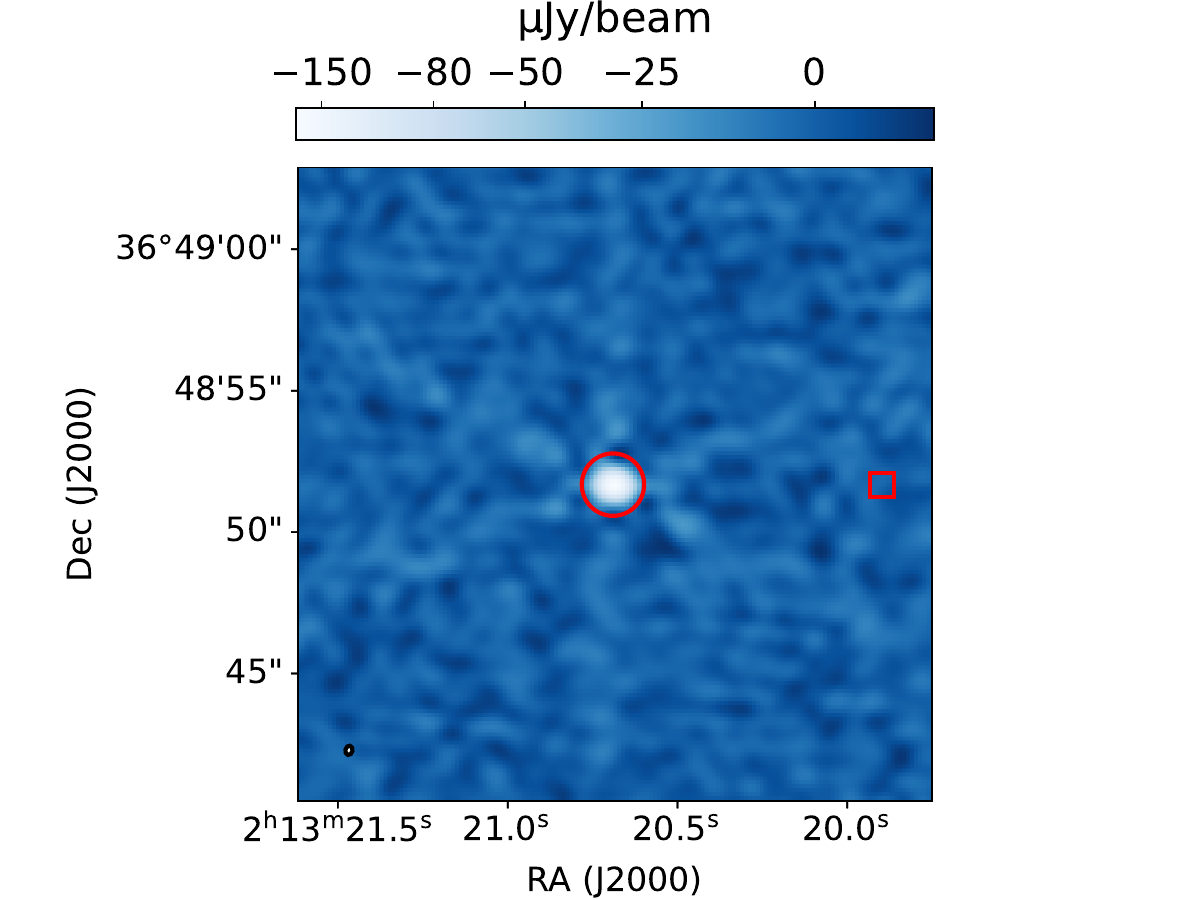} &
      \includegraphics[width=0.8 \columnwidth]{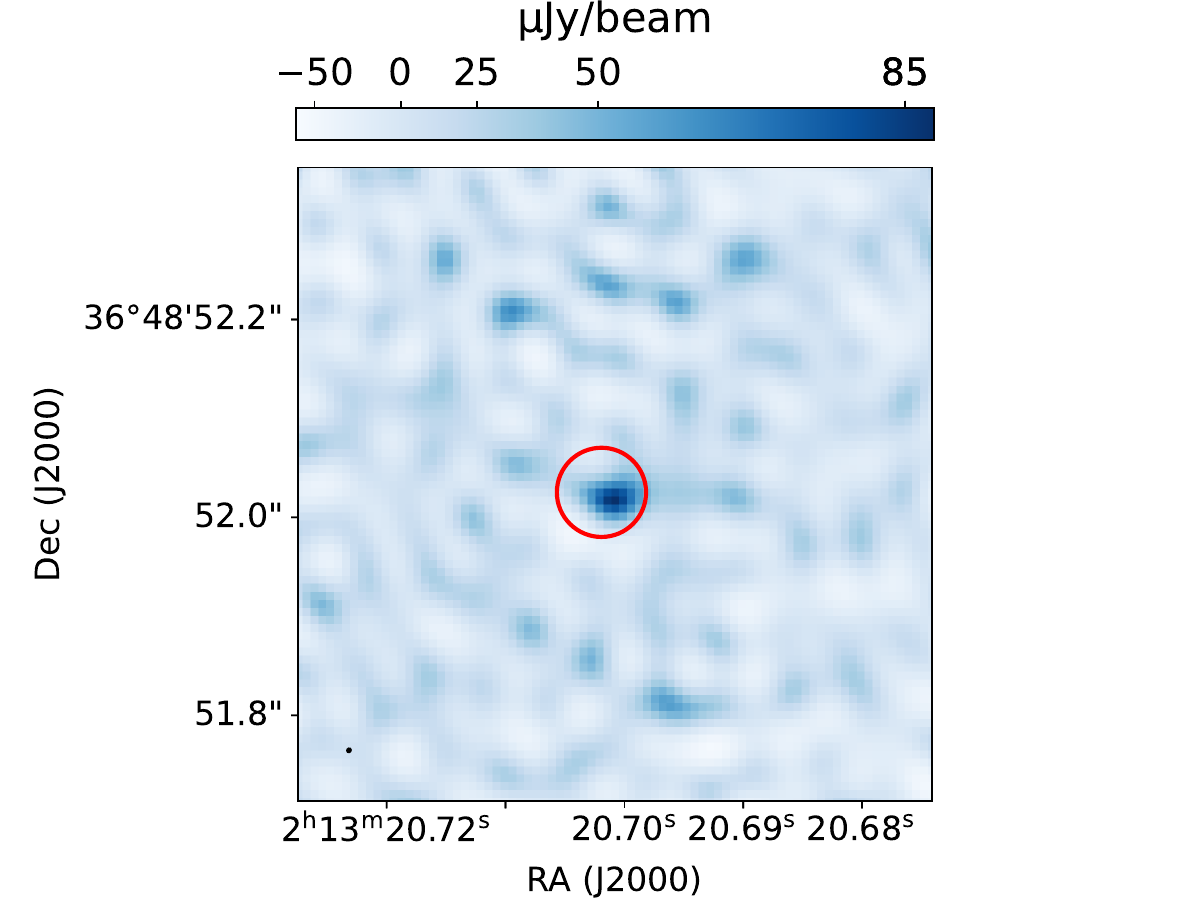} \\
      (a) & (b) & (c)
    \end{tabular}
    \caption{(a) Stokes I image of 2M0213 AB. The binary is unresolved and enclosed in the circle with a red outline. The binary is detected at a peak flux density of $\sim356\ \mu$Jy $\text{beam}^{-1}$. The $1\sigma$ r.m.s noise in the image is $\sim10\ \mu$Jy $\text{beam}^{-1}$ giving a SNR $\sim36$. The binary at the true coordinates at epoch 2017-11-15 is offset from the pointing centre by $\sim1.3$\arcsec, that is, by $\sim2$ synthesised beamwidths due to proper motion effects described in Section~\ref{section:astrometry}. The position of the wide T3 brown dwarf companion at a separation $\sim16.4$\arcsec is marked by the box. The red cross indicates the correlation centre ( pointing centre). (b) Stokes V image of 2M0213 AB. The binary is unresolved and enclosed in the circle with a red outline. The data has been phase-shifted to the true coordinates at epoch 2017-11-15.
 The binary is detected at a peak flux density of $\sim-174\ \mu$Jy $\text{beam}^{-1}$ where the negative indicates the left circular polarisation. The $1\sigma$ r.m.s noise in the image is $\sim4.6\ \mu$Jy $\text{beam}^{-1}$. The source is detected at SNR $\sim38$. Similarly, the position of the wide T3 brown dwarf companion is marked by the box. (c) Stokes I image of 2M0213 AB synthesised from e-MERLIN follow-up observations of the binary. The M4.5 component is detected and enclosed in a circle with a red outline. We do not show the position of the T3 dwarf since it is detected in the VLA data at a similar noise level to the e-MERLIN data at $\sim16\ \mu$Jy $\text{beam}^{-1}$. The synthesised beam in all images is indicated by the filled white circle with a black outline to the bottom left.}
 \label{fig:sources}
\end{figure*}

The pointing centre for the observations of 2M0213 ABC was based on coordinates from \cite{Janson2012} at epoch 2007-06-01. The scheduling did not account for the proper motion as the coordinates were not propagated to the VLA epoch of observation: 2017-11-15 which introduced an offset in the position of the system. For 2M0213 AB, we used the Gaia DR2 position \citep[][]{Gaia_DR2_astrometric2018} and proper motions from \citet[][]{Best2020} to obtain the true position of the binary at the time of observation. We compared the propagated position (02h13m19.8952 +36d48m51.6699s) to positions extracted from the the image (02h13m20.68396s +36d48m51.69034s) corresponding to an offset of $\sim31$ mas. The binary 2M0213 AB presented in Figure~\ref{fig:sources}(a) was detected at a peak flux density $356\pm6.1\ \mu$Jy $\text{beam}^{-1}$ at a signal-to-noise ratio (SNR) $\sim36$. The flux density and position were obtained by fitting a Gaussian using CASA's \textcolor{black}{\texttt{imfit} tool}. Considering the offset is less than the imaging cell size ($\sim140$ mas) we confirm the detection of the M dwarf binary. The wide T3 component at 16.4\arcsec is not detected, although we have marked its position on the image using a square box. We have used the same astrometric parameters to propagate the coordinates to the e-MERLIN epoch 2025-02-15. We have detected a component at SNR $\sim5$ (see Figure~\ref{fig:sources}(c)). Using a similar Gaussian fit, we have determined a peak flux density $90.5\pm7.5\ \mu$Jy $\text{beam}^{-1}$ at coordinates 02h13m20.70069s +36d48m52.02129s (at observing dates 2025-03-13 to 2025-03-14). We find an offset of $\sim12$ mas in position between the pointing centre coordinates from Gaia propagated to epoch 2025. We note the M6.5 component is in orbit at $\sim217$ mas, which is $\sim5$ synthesised beamwidths from the pointing centre.

\begin{figure*}
  \centering
    \setlength{\tabcolsep}{25pt} 
    \renewcommand{\arraystretch}{0} 
  \begin{tabular}{c}
      \includegraphics[width=0.8\columnwidth]{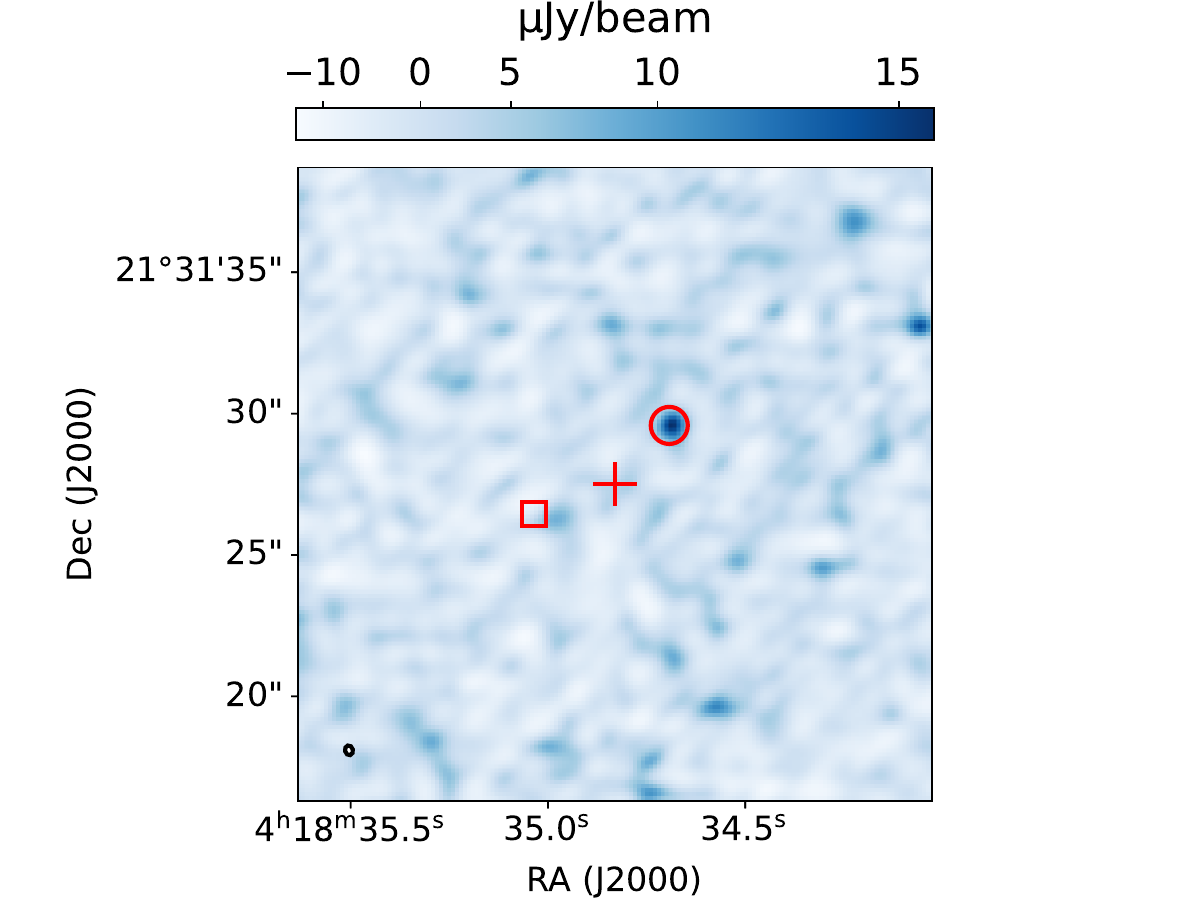} \\
    \end{tabular}
 \caption{Stokes I image of 2M0418. The position of the L5 brown dwarf is marked by the red square box in the image. A radio source was detected at $4\sigma$, where $\sigma$ is the thermal noise in the image, and at a peak flux density of $\sim16\ \mu$Jy $\text{beam}^{-1}$, and is enclosed in the red circle.
 The correlation centre (pointing centre) is indicated by the red-cross and the synthesised beam by the filled white circle with a black outline to the bottom left. We have not included Stokes V maps for observations of the L5 brown dwarf (2M0418) on account of the null detection.}
 \label{fig:2M0418}
\end{figure*}

The L5 brown dwarf 2M0418 in the Hyades cluster is too faint to be detected by Gaia. The pointing centre for the observation used coordinates from the original processing of 2MASS at epoch 1997-10-31. Fortuitously, \cite{Lodieu2019} have, however, used the Liverpool telescope in the infrared to constrain the proper motions at $\mu_\alpha\approx142.0\pm4.3$ $\text{mas\ yr}^{-1}$, $\mu_\delta\approx-51.8\pm4.0$ $\text{mas\ yr}^{-1}$ and the parallax at 25.8 ± 2.9 mas. The position at epoch 2015.4 is also available from the CATWISE catalogue, which is a reprocessing of the Wide-field Infrared Survey Explorer (WISE) and NEOWISE catalogues \citep{Marocco2021}. 
Propagating the coordinates reveals the brown dwarf has shifted by $\sim3$\arcsec. We have not made a detection at the true position  of 2M0418 indicated by a square box in Figure~\ref{fig:2M0418}.  We however detect a radio source $\sim5\farcs8$ from the true  position at $\sim4\sigma$ and at a peak flux density of $\sim16\ \mu$Jy $\text{beam}^{-1}$ . Using the source count flux density relationship $N(S) \simeq (23.2\pm2.8)S^{-1.18}$ arcmin$^2$ at 5 GHz \citep[][]{Fomalant1989} which places the density of sources at $>16\ \mu$Jy $\text{beam}^{-1}$ at $\sim2.45\times10^{-4}$ arcsec$^{-2}$, the probability of the source being a false positive in an area $\sim6\arcsec$ is $\sim2.7\times10^{-2}$.  Considering the large offset in position from the true position of 2M0418, the radio emission is unlikely to be the origin. However, it remains plausible that the emission is from a background AGN. There are no known associations between the source and existing radio catalogues.

\subsection{Variability}

\begin{figure*}
    \centering
    \includegraphics[width=0.9\linewidth]{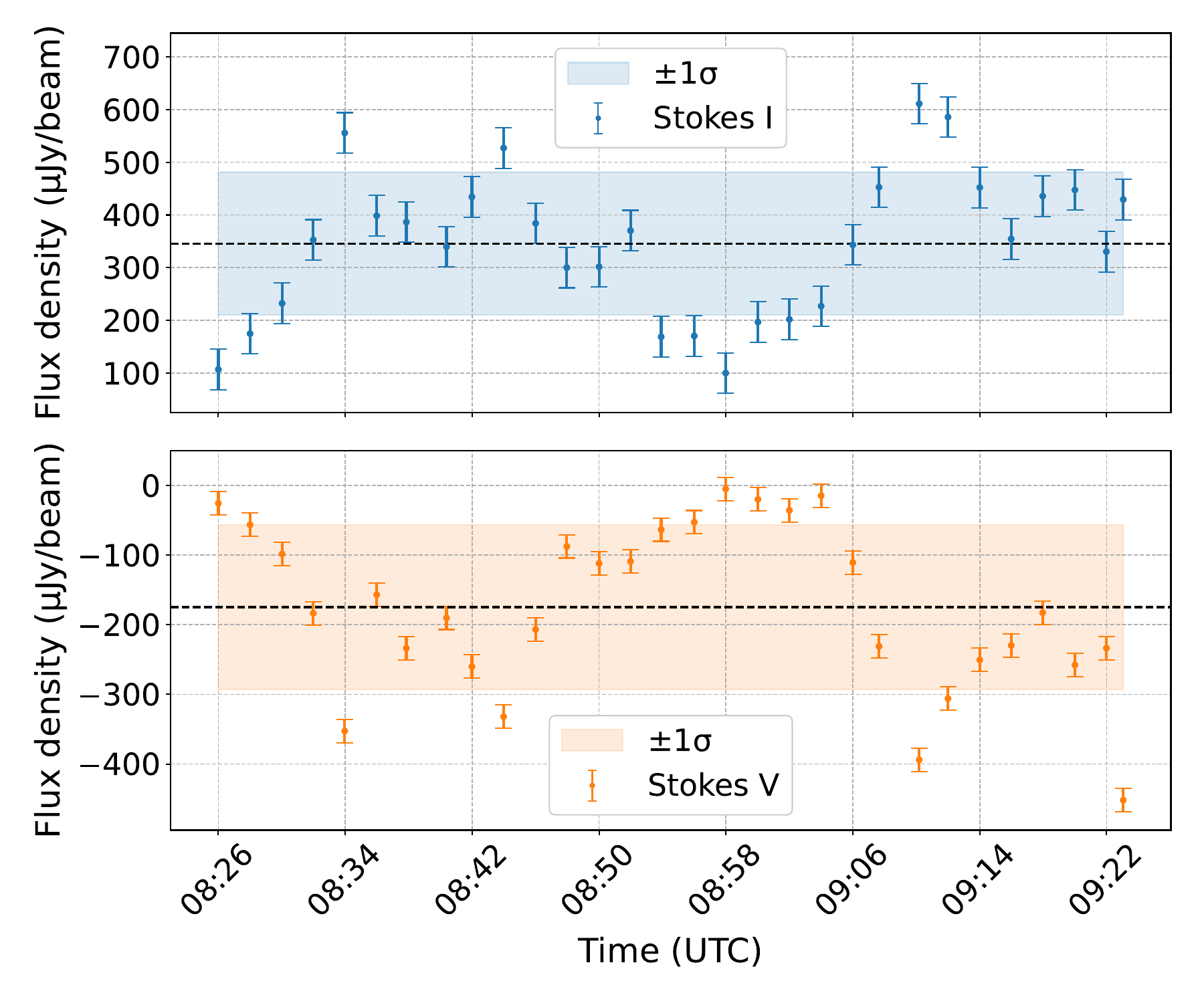}
    \caption{\textcolor{black}{Flux light curves of 2M0213AB from the VLA data binned to a cadence of two minutes. The observation was $\sim55$ minutes long. The upper and lower subplots show the Stokes I and V light curves respectively. The error bars represent the $1\sigma$ thermal noise added in quadrature to the flux scaling error and the shaded region represent a $1\sigma$ standard deviation of  
    $\sim133\ \mu\text{Jybeam}^{-1}$ and $\sim118\ \mu\text{Jybeam}^{-1}$ in the Stokes I and V respectively. The mean flux densities in the Stokes I and V are $\sim347\ \mu\text{Jybeam}^{-1}$ and $\sim-176\ \mu\text{Jybeam}^{-1}$ respectively and are indicated by the dashed lines in the respective subplots. Note: These values slightly differ from values obtained using \texttt{imfit}}.}
    \label{fig:2M0213_lightcurve}
\end{figure*}

\begin{figure}
    \centering
    \includegraphics[width=\linewidth]{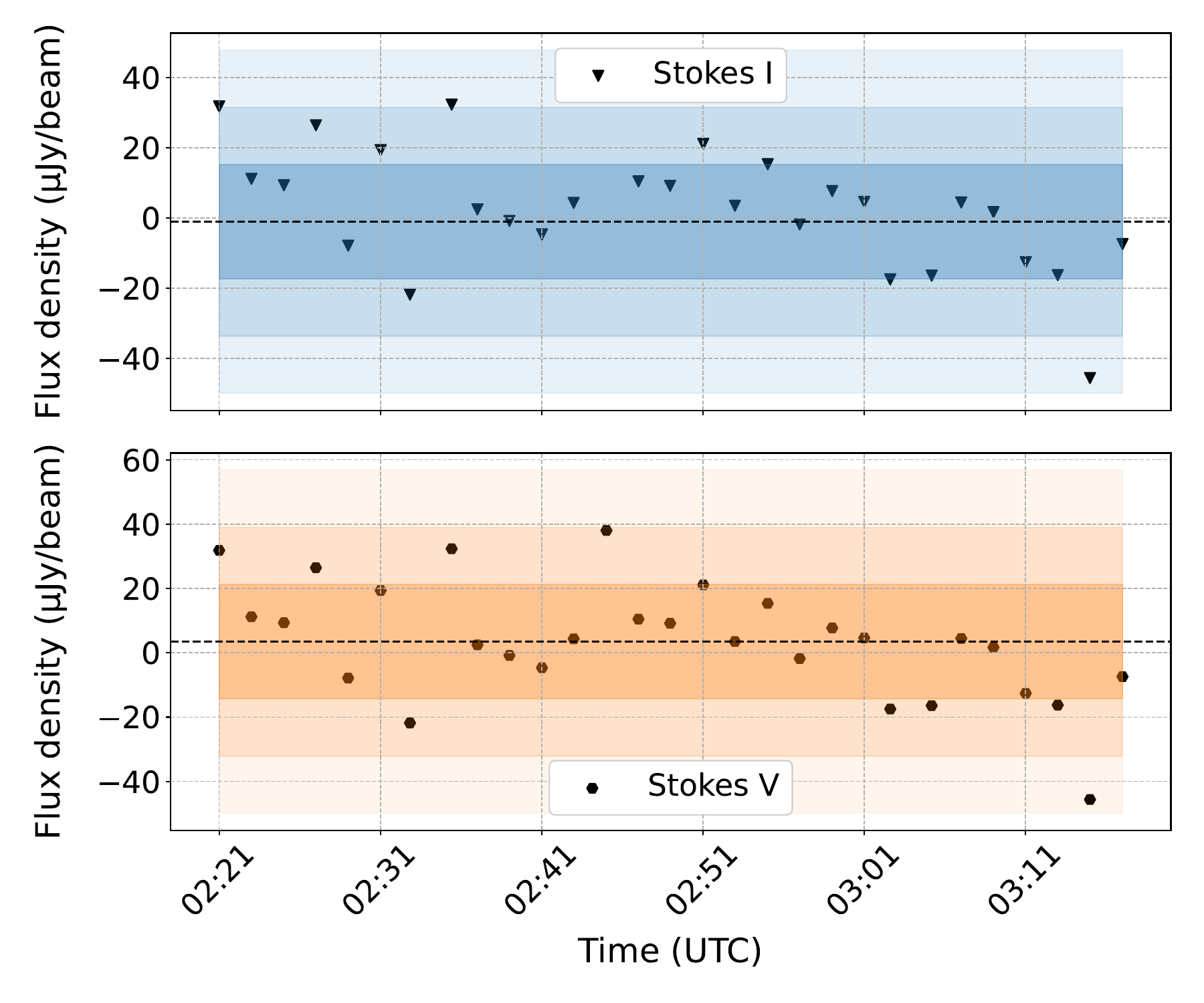}
    \caption{\textcolor{black}{Flux light curves of 2M0418 from the VLA data binned to a cadence of two minutes. The observation was $\sim55$ minutes long. The upper and lower subplots show the Stokes I and V light curves respectively. the shaded region indicate a 1,2 and 3$\sigma$ standard deviations from the mean flux density of $\sim-1.0\ \mu\text{Jybeam}^{-1}$ and $3.5\ \mu\text{Jybeam}^{-1}$ for the Stokes I and V is respectively. The mean values are represented by the dashed lines. The $1\sigma$ standard deviation for the Stokes I and V is $\sim16.3\ \mu\text{Jybeam}^{-1}$ and $\sim17.8\ \mu\text{Jybeam}^{-1}$ respectively which is consistent with the VLA's $1\sigma$ thermal noise estimated for a two minute cadence. }}
    \label{fig:2M0418_lightcurve}
\end{figure}

\textcolor{black}{To probe for variability, we first bin the visibilities at a two minute cadence as a compromise between sensitivity and temporal resolution and proceed to produce Stokes I and V light curves from the \texttt{}{RR} and \texttt{LL} correlations. We obtain the uncertainties by adding the thermal noise for a dual polarised robust weighted image ($\sigma_\text{t}\sim16.7\ \mu\text{Jybeam}^{-1}$) observed over a cadence of two minutes \footnote{\url{https://obs.vla.nrao.edu/ect/}} to a flux scaling error of $\sim10$\% in quadrature. Flux scaling errors are associated with the flux scaling calibrators and occur due to challenges in determining the absolute flux density. The adopted flux scaling error is consistent with the recommendations provided by the VLA \footnote{\url{https://www.vla.nrao.edu/astro/calib/vlacal/cal_mon/last/1331+3030.html}}.  It should be noted that we include the flux scaling error for the Stokes I only as the Stokes V errors are uncharacterised.  
From the light curves generated from observations of 2M0213 AB presented in  Figure~\ref{fig:2M0213_lightcurve}, we do not detect any statistically significant deviations at $>3\sigma$ from the mean indicating the source is in a quiescent state.
Following the non detection of 2M0418, we make a light curve binned at the same cadence as the 2M0213 AB light curve  and estimate the uncertainties at a $3\sigma_\text{t}$ level. No pulses are detected as the flux densities are consistent with noise at a $3\sigma_\text{t}$ level. We present the Stokes I and V light curve of 2M0418 in Figure~\ref{fig:2M0418_lightcurve}.  }

\subsection{The Spectral Energy Distribution (SED) of the Binary}
\label{section:sed}

\begin{figure}
    \centering
    \includegraphics[width=\linewidth]{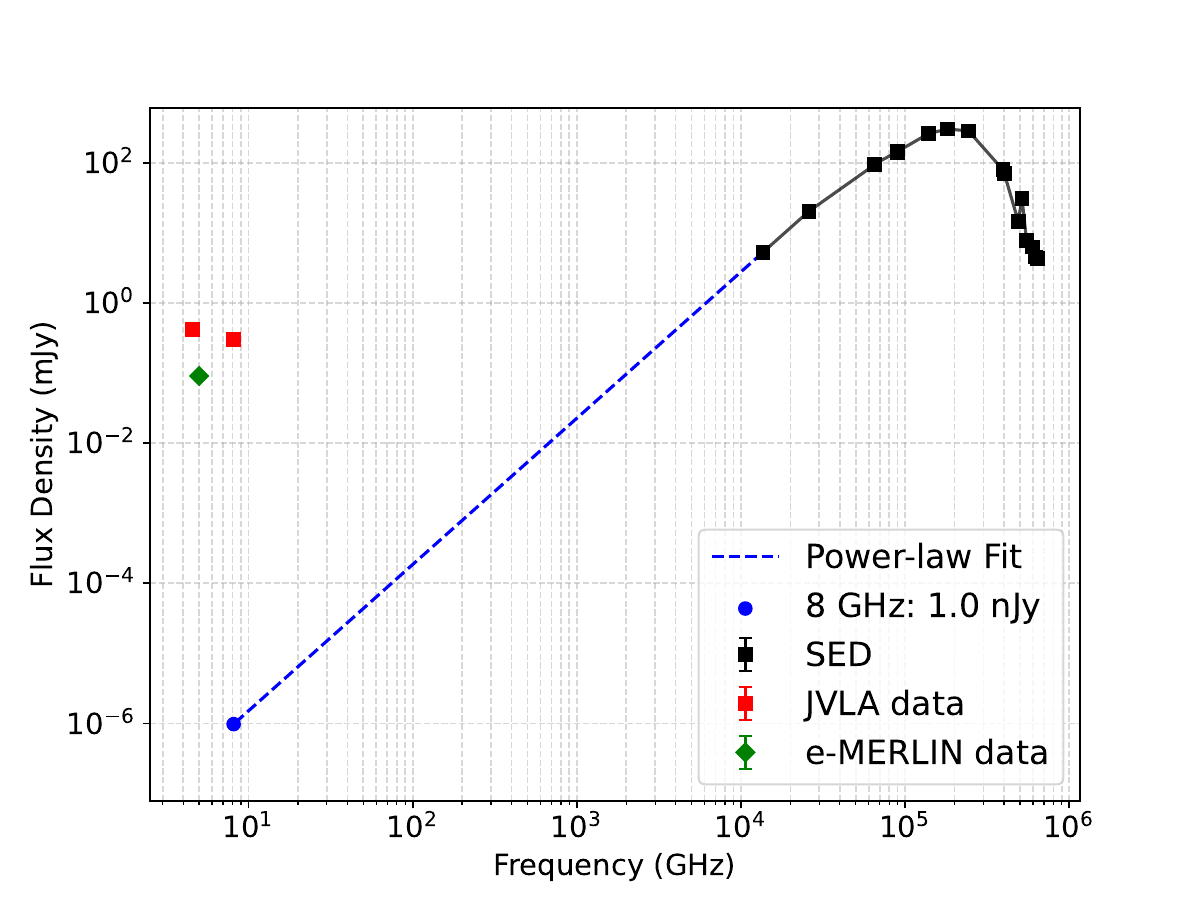}
    \caption{An SED fit of the unresolved binary. The black curve represents the SED fit using different photometric catalogues described in Section~\ref{section:sed}, with each data point representing a different catalogue. Flux error bars are included for each data point, although they are not visible due to the scaling. The two red data points depict VLA fluxes at 4 and 8 GHz, respectively. The green data point depicts e-MERLIN flux at 5 GHz. The blue broken line linking the SED data points to the blue data point at 8 GHz is a power law fit, which represents the expected flux for an SED extrapolated to 8 GHz.
    }
    \label{fig:sed_fit}
\end{figure}

For the SED fitting, we used {\sc PySSED} \citep[][]{pyssed2024}, a Python tool for quick SED fitting. This tool collected optical photometry from the Panoramic Survey Telescope \& Rapid Response System Data Release 1  \citep[Pan-STARRS;][]{Panstarrs2016} in the \textit{grizy} filters, the American Association of Variable Star Observers's
Photometric All-Sky Survey \citep[APASS;][]{APASS2014} using the Johnson $V$ and Sloan Digital Sky Survey (SDSS) $gri$ filters, Gaia \citep[][]{Riello2021,Gaia_summary_2023} $B_P,G,R_P$ filters, the Carlsberg Meridian Catalog 15 \citep[][]{CM152014} using SDSS $r$ filter, 2MASS \citep[][]{Cutri2003} using the $VJHK$ filters, the Wide-field Infrared Survey Explorer \citep[WISE][]{WISE2010} in the mid-infrared at bandpasses of 3.4, 4.6, 12 and 22 $\mu$m and the reprocessed versions of WISE, catWISE \citep[][]{Marocco2021} and unWISE \citep[][]{Lang2014}. The {\sc bt-settl} AGSS2009 atmosphere models \citep[][]{Allard-2011} have been used to generate model atmospheres. {\sc PySSED}'s default fitting parameters were used. The fit returns an effective temperature $T_\text{eff}\sim3016$ K and a stellar radius of $R_\star\sim0.25R_\odot$. These values are characteristic of a mid-M dwarf and are consistent with the values for a M4\,V dwarf derived by \cite{Pecaut2013} with $T_\text{eff}\sim3200$ K. We note that the binary is unresolved in these catalogues. We have included the VLA measured flux densities at \textcolor{black}{the edges of the lower and upper end of the observing band i.e.,} 4 and 8 GHz \textcolor{black}{and are shown} in the SED fit in Figure~\ref{fig:sed_fit}. We have further extrapolated the fit to 8 GHz by fitting a simple power law under the assumption that the fluxes follow the trend of a concomitant decrease of the flux density with decreasing frequency. We note that although this idea is basic and a break in the power law is expected, it illustrates that the radio emission, which has flux densities $\sim10^5$ larger than predicted by the power law, is generated by a very powerful mechanism presumably powered by magnetic activity.

\subsection{Nature of the Radio Emission}

To characterise the nature of the radio emission, we first measure the spectral index and the circular polarisation fraction using the VLA data due to it's wide observing bandwidth, estimate brightness temperatures and luminosities, discuss the discrepancy in the flux densities measured using the VLA and e-MERLIN, and finally constrain the emission mechanism. 

\subsubsection{Spectral Index and Polarisation}
\label{section:spectral_index}

Following the procedure for measuring the spectral index described in section~\ref{section:data_analysis}, we determine an almost flat spectral index ($S\propto\nu^{\alpha}$) at \textcolor{black}{$\alpha=-0.44\pm0.07$ }. \textcolor{black}{We estimate the} circular polarisation fraction \textcolor{black}{$f_\text{c}$}  by imaging the VLA data in the Stokes I and V (see Figures~\ref{fig:sources}(a) and (b)) and evaluating the ratio $f_\text{c}=|V|/I$. The measured flux densities in Stokes I and V are $356\pm6.1\ \mu$Jy $\text{beam}^{-1}$ and $-179.2\pm3.0\ \mu$Jy $\text{beam}^{-1}$ respectively. Using the flux densities, \textcolor{black}{we determine } $\textcolor{black}{f_\text{c}}=50.3\pm1.2$\% and a mean $\textcolor{black}{f_\text{c}=45.20 \pm 1.58}$\% with the data binned to a two minute cadence.

\subsubsection{Brightness Temperature and Luminosity}

We determine the brightness temperature,  $T_\text{B}$ of the emission using the following relation \citep[see][]{Dulk1985,Burgasser2005}
\begin{equation}
    T_\text{B} \simeq \frac{S_\nu}{\text{1 mJy}}\times\ \left(\frac{\nu}{\text{1 GHz}}\right)^{-2} \times \left(\frac{d}{\text{1 pc}}\right)^2\ \times \left(\frac{L}{\text{1 cm}}\right)^{-2} \times10^{29}\ \ \  \text{K} \ ,
    \label{eqn:brightness_temp}
\end{equation}
where $S_\nu$ is the flux density in mJy, $\nu$ is the frequency in GHz, $d$ is the distance to the binary in parsecs and $L$ is the length of the emitting region in cm. We highlight that although we have made detections using the VLA and e-MERLIN data, useful brightness temperatures cannot be extracted from the images due to their low resolutions (1\arcsec at 6 GHz in the VLA's B configuration and 50 mas at 5 GHz for e-MERLIN).  Following \cite{Berger2002,Bao2007} we assume $L$ $\sim0.1$ to $1R_\star$ where $R_\star$ is the radius of the M4.5 primary. By adopting this approach and restricting the emission to a small region, we use equation~\ref{eqn:brightness_temp} assess whether the brightness temperatures are in excess of the upper limit for incoherent emission from stellar corona at  $\sim10^{10}$ K \citep[][]{Dulk1985}. We determine $T_\text{B}\sim6.6\times10^{10}$ K for $L\sim0.1R_\star$ and $T_\text{B}\sim6.6\times10^{8}$ K for $L\sim1R_\star$.  We highlight that our brightness temperature estimates neglect the source's morphology.

To determine the electron energies, we begin the analysis by assuming  the source is optically thin ($\tau_\nu\ll1$) and isolated and has a brightness temperature $T_\text{B}$. The effective temperature of the electrons $T_\text{eff}$ is then determined as $T_\text{B}=\tau_\nu\ T_\text{eff}$ and for an optically thick source $T_\text{B}=T_\text{eff}$ \citep[][]{Dulk1985}. The lower bound for $T_\text{eff}$ must then be $T_\text{eff} \geq  T_\text{B}$ which is also consistent with an optically thick source. Admittedly $T_\text{eff}$ can be determined using the models of \cite{Dulk_Marsh1982} although they are restricted in applicability and require knowledge of the magnetic field strength $B$.

Using the brightness temperatures, we estimate electron energies, $E\sim56.52$ keV to 5.65 MeV for $T_\text{eff}\sim6.6\times10^8-6.6\times10^{10}$ K. The effective temperature in the MeV results from the small length scale used. Finally, we determine a bolometric luminosity $\text{log}_{10}\ L_\text{bol}\sim31.25\ \text{erg}\ \text{s}^{-1}$, a spectral luminosity ($L_\nu=4\pi S_\nu d^2$) of $\text{log}\ L_\nu \sim13.94\ \text{erg}\ \text{s}^{-1}\text{Hz}^{-1}$ and a radio luminosity  $\left(L_\text{R}\approx L_\nu\Delta\nu\right)$ of $\text{log}_{10} L_\text{R}\sim23.54\ \text{erg}\ \text{s}^{-1}$. $L_\text{bol}$ is determined from the Stefan Boltzmann's law using the effective temperature obtained from the SED fitting.
The  radio luminosity to the bolometric luminosity  is $\text{log}_{10}L_R/\text{log}_{10}L_\text{bol}\sim-7.76$. We note that the luminosities determined from the VLA observations are upper limits. A summary of the values obtained from the analysis is presented in Table~\ref{table:analysis_summary}.

\subsubsection{\textcolor{black}{VLA and e-MERLIN Flux Discrepancy}}

\textcolor{black}{The measured flux density using the VLA at $\sim356\ \mu\text{Jybeam}^{-1}$ is $\sim4$ times greater than the corresponding e-MERLIN flux density. This difference may arise from a combination of factors, two of which are examined in detail here. The first factor to consider is the intrinsic variability of the binary. M dwarfs have been observed to exhibit long term magnetic cycles, for example, \citet[][]{Buestos2025} have recently measured cycling periods ranging from 3 to 19 years using 13 stars of spectral types ranging M0 to M6. Proxima Centauri, the mid-to-late M dwarf of spectral type M5.5 has a measured cyclic activity occurring at a period ranging from $\sim442$ days \citep[][]{Cincu2007} to 7-8 years \citep[][]{Wargelin2017,Wargelin2024} and brightens in the X-rays by a factor of $\sim1.5$ at stellar maximum \citep[][]{Wargelin2024}. 
Conversely, short term variability is routinely observed in M dwarfs e.g. observations of Proxima Centauri between 1.1 to 3.1 GHz have detected intra-day variability attributed to flaring activity. The variability peaks at flux densities of 25 and 45 mJy which exceeds the average flux density by factors of 100 and 200 respectively \citep[see][]{PerrezTorres2021}. Consequently, considering the VLA observations are separated from the e-MERLIN observations by $\sim8$ years and the finding that mid-to-late M-dwarfs display short term variability, we cannot rule out intrinsic variability as a contributing factor to the flux difference.} 

\textcolor{black}{Secondly, the VLA and e-MERLIN probe different scales owing to the the baseline lengths; the VLA in its B configuration has a maximum baseline of  11 km and e-MERLIN 217 km resulting in different synthesised beam ($\theta_\text{synth}$) sizes. B configuration VLA observations at 6 GHz result in  $\theta_\text{synth}\sim1$\arcsec and e-MERLIN observations at 5 GHz result in  $\theta_\text{synth}\sim40$ mas. Accordingly, for the binary at a distance $\sim14.28$ pc, different linear spaces are probed. The VLA probes a linear scale of $\sim14.28$ AU while the e-MERLIN probes a linear scale of $\sim0.57$ AU. Therefore spatial filtering due to the disparate synthesised beam sizes cannot be dismissed as the higher resolution e-MERLIN could resolve out large scale emission that is otherwise well sampled by the low resolution VLA. Assuming a stellar radii, $R_\star\sim 0.25\ R_\odot$ (see section~\ref{section:sed} for the justification), the spatial scale probed by e-MERLIN is of size $\sim490\ R_\star$. Similarly to \citet[][]{Climent2022}, we find such a source size improbable, which seemingly implies the detected radio emission should originates from both components. However, this interpretation conflicts with the 
separation distance of $\sim$0.217\arcsec \citep[3.1 AU, ][]{Janson2014}. Although the two components are within the synthesised beam of the VLA, they are separated by $\sim5$ beamwidths in the e-MERLIN observation.  With an orbital period of $6.13-7.15$ years \citep[][]{Janson2014}, the binary separation distance is unlikely to significantly evolve over the timescale of the VLA and e-MERLIN observing sessions to $\lesssim40$ mas. Consequently, we attribute the flux difference to variability. We recommend both higher resolution observations of the system and continued monitoring to characterise the variability.}

\begin{table}
    \renewcommand{\arraystretch}{1.4}
    \centering
    \begin{tabularx}{\columnwidth}{ll} 
		\hline
		Parameter & Value \\
        \hline
        Spectral Index & \textcolor{black}{$-0.44\pm0.07$} \\
        Brightness temperature, $T_\text{B}$ & $<10^{11}$ K \\
        Bolometric luminosity, $\text{log}_{10}\ L_\text{bol}$ & 31.25 $\text{erg s}^{-1}$ \\
        Radio luminosity, $\text{log}_{10}\ L_\text{R}$ & 23.54 $\text{erg s}^{-1}$ \\
        Mean circular polarisation fraction, \textcolor{black}{$f_\text{c}$} & \textcolor{black}{$45.20 \pm 1.58$\%} \\
         \hline
	\end{tabularx}
    \caption{Summary of results from the analysis}
    \label{table:analysis_summary}
\end{table}

\subsubsection{Emission Mechanism}

A plausible radiation mechanism for the underlying spectral characteristics is  mildly relativistic electrons radiating gyrosynchrotron emission. Following the spectral turnover at $\alpha=0$, the emission is largely characterised by an optically thin spectral index and is produced by mildly relativistic electrons following a power law distribution $n (E) \propto E^{-\delta}$ where $\delta$ is the power law index. In this regime the emission exhibits a spectral index $\alpha=1.22-0.9\delta$ \citep[see][]{Dulk_Marsh1982} and a spectral power $\nu^{5/2}$ in the optically thick case \citep[see][]{Gudel2002}. \citet[][]{Gudel2002} have demonstrated that $2\lesssim\delta\lesssim4$ in stellar corona. We determine \textcolor{black}{$\delta\sim1.84$} which is consistent with their findings. The mean circular polarisation fraction is on the high end of gyrosynchrotron emission. \textcolor{black}{However, \cite{Golay2023} demonstrated that a gyrosynchrotron emission, which occurs at harmonics between 10 and 100 times the electron gyrofrequency, from a uniform magnetic field at optical depths $\ll$ 1 exhibits fractional polarisation ratios ranging from $|V|/I\sim90\%$ at lower harmonics to 10\% at higher harmonics.} We highlight that the flux densities from the Stokes V image are negative indicating the left circular polarisation is dominant (see figure~\ref{fig:sources}). In such a case, the emission is polarised in the sense of the ordinary (O)-mode suggesting optically thick gyrosynchrotron emission \citep[see][]{Dulk1979} seemingly presenting a  contradiction with the interpretation of the spectral index. We note high polarisation fractions are not uncharacteristic of optically thin gyrosynchrotron emission with pitch angle distributions. In such a scenario, the polarisation mode is dependent on the viewing angle and the shape of the distribution \citep[see][]{Fleishman2003}.

\textcolor{black}{To estimate the magnetic field strength $B$, the length of the emitting region $L$ and the electron number density $n_\text{e}$  we apply the semi empirical  solutions of the radiative transfer equation for gyrosynchrotron radiation from a non-thermal distribution of electrons by \citet[][]{Dulk1985,Gudel2002}. We assume a viewing angle $\theta\sim\pi/3$ and $\delta=3$ which is typical for stellar corona and similarly to \citet[][]{Berger2006}, we determine the peak frequency $\nu_\text{p}$ as 
\color{black}
\begin{equation}
    \nu_\text{p} \approx 16.6 \times n_\text{e}^{0.23} L^{0.23}\times B^{0.77} \times10^{3}\ \ \text{Hz},
\end{equation}
the flux density as
\color{black}
\begin{equation}
S_{\nu,\text{p}} \approx1.54 \times B^{-0.76} \times L^2 \times d^{-2} \times \nu_\text{p}^{2.76} \times 10^{-4} \ \ \mu\text{Jy}, 
\end{equation}
\color{black}
and the fractional polarisation ratio as
\begin{equation}
f_\text{c} \approx2.85\times B^{0.51} \times\nu_\text{p}^{-0.51}\times 10^{3}\ \ .
\end{equation}
}

\textcolor{black}{Using $f_\text{c}\sim50.3$\% which is estimated from the image plane and assuming the emission  peaks at a frequency (at which the plasma transitions from optically thick to optically thick) $v_\text{p}\lesssim4$ GHz, we estimate $B<174.86$ G, $L<1.54R_\star$ and  $n_\text{e}<2.91\times10^{5}\ \text{cm}^{-3}$.}

We cannot entirely dismiss coherent emission produced through the electron cyclotron maser emission mechanism (ECME) due to the brightness temperatures which are in excess of $10^{10}$ for $L=0.1R_\star$. For ECME the radio emission is at a local cyclotron $ \nu_c = 2.8\times10^{-3}\ B$ GHz which constrains the stellar magnetic field strength $B\sim1.4-2.8$ kG.

Based on the spectral index,circular polarization fractions, which, although high, do not reach 100\% fractional polarization typical of ECME \citep[][]{Hallinan2008},  the calculated luminosities which are consistent to the same order of magnitude with luminosities for quiescent emission from M dwarfs \cite[e.g.][]{Burgasser2005}, lower bounds of electron energies which are in the range $1\sim100$ keV and brightness temperatures (assuming an emission region on the scale of the stellar disk) and the \textcolor{black}{persistent emission}, we argue the emission is consistent with gyrosynchrotron radiation.

\section{Conclusions}

We have detected 2M0213 AB, a binary M dwarf system  \textcolor{black}{at} a peak flux density of $\sim356\ \mu$Jy beam$^{-1}$ using the VLA. By employing various astrometric catalogues, we have conclusively determined that the detected source corresponds to the binary. The radio emission from the binary is polarised at a mean circular polarisation fraction $f_\text{c}=45.2\pm1.58$\% and exhibits a spectral index \textcolor{black}{$\alpha=-0.44\pm0.07$ }. We have used photometric SED fitting to  constrain the radius of the M4.5 component to $0.25M_\odot$ and the effective temperature to 3016 K. The radio luminosity of the M dwarf binary is $\text{log}L_R/\text{log}L_\text{bol}\approx-7.76$. We have made follow-up observations of the binary using the e-MERLIN and detected a single component. From our higher angular resolution detection, we argue the emission at 5 GHz is \textcolor{black}{potentially associated with the M4.5 primary, which may also display short or long term variability.  Using Stokes I and V light curves, we have searched for short-term variability over the duration of the VLA observation and failed to detect any emission at $>3\sigma$ from the mean indicating the emission is quiescent. Continued monitoring of the binary using very long baseline interferometry should conclusively determine the active component and constrain the true variability.} Based on the electron energies, the brightness temperatures, radio luminosities and the spectral index, we have demonstrated the emission is consistent with gyrosynchrotron radiation. It is noteworthy that this detection adds to the already rare catalogue of binaries that have been probed for radio emission. \textcolor{black}{\cite{Kao2025} have recently shown that binarity in UCDs increases the occurrence of radiation belts which are a prevailing explanation for the origin of quiescent emission in UCDs \citep[see][]{Leto2021,Climent2022}. As such continued monitoring of binary systems not only adds to the sparse catalogue of radio active M dwarfs but may also determine the origin of the emissions.}

\textcolor{black}{ Regrettably, we did not detect 2M0418. We detected a background AGN at an angular distance of $\sim5.8\arcsec$ from the position of 2M0418 and at a peak flux density $\sim16\ \mu$Jy $\text{beam}^{-1}$ . We have produced a light curve for the undetected 2M0418, searched for short duration bursts and made no detection at a significant level $>3\sigma$ from the mean. We note that among UCDs, L dwarfs have the lowest radio activity as $\sim5$\% exhibit flares \citep[][]{Route2016_arecibo_ucd_stats} and $\sim10-13$\%  display quiescent emission \citep[][]{Kao2024}. 
We highlight that chromospheric activity becomes increasingly rare towards later L spectral types \citep[][]{Schmidt2015} with $\sim9.3$\% of mid-to-late L dwarfs exhibiting chromospheric activity in the form of H$\alpha$ emission \citep[][]{Pineda2016}. As such, 
the non detection of 2M0418 is unsurprising considering the low detection fractions for L dwarfs. 2M0418 nevertheless remains intriguing given its chromospheric activity which is rare in mid L dwarfs. }

\section{Acknowledgements}

This project has been made possible in part by a grant from the SETI Institute. This work made
use of Astropy:3, a community-developed core Python package and an
ecosystem of tools and resources for astronomy \citep[][]{astropy2013,astropy2018,astropy2022}. This work has made use of data from the European Space Agency (ESA)
mission {\it Gaia} (\url{https://www.cosmos.esa.int/gaia}), processed by
the {\it Gaia} Data Processing and Analysis Consortium (DPAC,
\url{https://www.cosmos.esa.int/web/gaia/dpac/consortium}). Funding
for the DPAC has been provided by national institutions, in particular,
the institutions participating in the {\it Gaia} Multilateral Agreement. The National Radio Astronomy Observatory is a facility of the National Science Foundation
operated under cooperative agreement by Associated Universities,
Inc. This research has made use of the SIMBAD database, operated at CDS, Strasbourg, France. This research has made use of the VizieR catalogue access tool, CDS, Strasbourg, France. e-MERLIN is a National Facility operated by the University of Manchester at Jodrell Bank Observatory on behalf of STFC. This publication makes use of data products from the Wide-Field Infrared Survey Explorer, which is a joint project of the University of California, Los Angeles, and the Jet Propulsion Laboratory/California Institute of Technology, funded by the National Aeronautics and Space Administration. We thank the anonymous referee for useful suggestions that have significantly improved the manuscript.

\section*{Data Availability}
Data underlying this article are publicly available in the NRAO Data Archive at \url{https://data.nrao.edu/portal} and can be accessed with project code 17B-169. The e-MERLIN data will be provided upon reasonable request.


\bibliographystyle{mnras}
\bibliography{mnras_template}







\bsp	
\label{lastpage}
\end{document}